\documentclass[11pt]{article}
\usepackage[margin=1in]{geometry}
\usepackage{graphicx}
\usepackage{url}
\usepackage{authblk}
\usepackage{amsmath}
\usepackage{enumitem}
\usepackage{natbib}
\usepackage{hyperref}
\hypersetup{colorlinks=true, linkcolor=blue, citecolor=blue, urlcolor=blue}

\title{Evolution of A4L: A Data Architecture for AI-Augmented Learning}
\author[1,2]{Ploy Thajchayapong}
\author[3]{Suzanne Carbonaro}
\author[3]{Tim Couper}
\author[3]{Blaine Helmick}
\author[1,2]{Spencer Rugaber}
\author[1,2]{Ashok Goel}
\affil[1]{National AI Institute for Adult Learning and Online Education}
\affil[2]{Design Intelligence Laboratory, Georgia Institute of Technology}
\affil[3]{1EdTech}
\date{}

\begin{document}
\maketitle

\begin{abstract}
As artificial intelligence (AI) becomes more deeply integrated into educational ecosystems, the demand for scalable solutions that enable personalized learning continues to grow. These architectures must support continuous data flows that power personalized learning and access to meaningful insights to advance learner success at scale. At the National AI Institute for Adult Learning and Online Education (AI-ALOE), we have developed an Architecture for AI-Augmented Learning (A4L) to support analysis and personalization of online education for adult learners. A4L1.0, an early implementation by Georgia Tech’s Design Intelligence Laboratory, demonstrated how the architecture supports analysis of meso- and micro-learning by integrating data from Learning Management Systems (LMS) and AI tools. These pilot studies informed the design of A4L2.0. In this chapter, we describe A4L2.0 that leverages 1EdTech Consortium’s open standards—such as Edu-API, Caliper Analytics\textsuperscript{\textregistered}, and Learning Tools Interoperability\textsuperscript{\textregistered} (LTI)—to enable secure, interoperable data integration across data systems like Student Information Systems (SIS), LMS, and AI tools. The A4L2.0 data pipeline includes modules for data ingestion, preprocessing, organization, analytics, and visualization. 
\end{abstract}


\section{Introduction}
Artificial Intelligence (AI) is offering unprecedented opportunities for personalization and scalability, and thereby rapidly reshaping the landscape of education. For adult learners in particular—who often face barriers related to age, health, family, and work —online education powered by AI presents a transformative model of accessible and adaptive instruction. However, the promise of AI-augmented learning can only be fully realized when supported by a robust, connected data infrastructure: personalization of learning depends on continuous, granular, and secure access to learner data that can inform near real-time instructional decisions by both human instructors and AI agents.

The problem of access to meaningful learner data is compounded by the segmentation of learner data in educational institutions. At many institutions of higher education, learner data is distributed across multiple, and often disparate, systems such as Enterprise Data Management (EDM), Student Information Systems (SIS), and Learning Management Systems (LMS). Furthermore, the various data systems typically are isolated from one another, use different formats, and are managed by different administrative units, making it all but impossible to use data from more than one system. This problem is especially acute for adult learners because of accumulation of data at multiple educational institutions across a lifetime.   

To address this need, the National AI Institute for Adult Learning and Online Education (AI-ALOE) has developed the Architecture for AI-Augmented Learning (A4L)—a scalable data pipeline tailored to online adult education. In earlier work, we described the motivations and goals for A4L (\cite{goel2024a4l}). We presented ecosystems for adult learning and online education that build in part on  \cite{lyndgaard2024lifelong} framework of micro-learning (moment-to-moment interactions with learning tools), meso-learning (learning patterns across a course), and macro-learning (long-term educational progress).  We also described the first instantiation of A4L (A4L1.0) developed by Georgia Tech’s Design Intelligence Laboratory as well as initial implementations of A4L1.0 for analyzing and visualizing data from learning tools such as the virtual teaching assistant named Jill Watson (\cite{kakar2024a}).

While A4L1.0 supports learning analytics on LMS, EDM, and AI tool data through a manually operated pipeline, it lacks integration with Student Information System (SIS) data. Furthermore, it requires labor-intensive manual processes to move and prepare data for analysis since aggregation across tools happens at this layer. While it enables analyses of micro- and meso-learning analyses, it faces scalability limitations due to a lack of automated ingestion, anonymization and standardization of data, especially across tools. 

Subsequently, the A4L team developed new design guidelines for A4L2.0 based on the initial experiments with A4L1.0. We then redesigned the data pipeline and evolved A4L1.0 toward a more streamlined and integrated architecture that we call A4L2.0. A4L2.0 introduces a fully asynchronous data ingestion model using 1EdTech Consortium’s open standards such as Edu-API \footnote {1EdTech https://www.1edtech.org/standards/edu-api}, Caliper Analytics\textsuperscript{®} \footnote{1EdTech https://www.1edtech.org/standards/caliper} and LTI\textsuperscript{®} \footnote{1EdTech https://www.1edtech.org/standards/lti} . The new plan was designed to accommodate A4L operating requirements - to deliver a testable and repeatable, software-defined infrastructure, and to ensure transparency and change management through existing established tools and processes like Pull Requests. The architecture is pluggable, so it imposes no dependencies on specific data or cloud infrastructure, allowing the design to be scaled from simple, cost-effective relation databases through much-larger-scale data lake environments if needed in the future. The data design is driven by open standards, not by specific vendor services, so provides autonomy and flexibility to future development of A4L. The design incorporates key components including automated endpoints, JSON schema validation, event triggers, queues, and parallel task execution for data transformation, anonymization, and aggregation. These enhancements fully integrate SIS data, automate analytics through time-based scheduling, and support privacy-preserving visualizations for different stakeholders using a data governance model and shifted data aggregation to the point of ingestion. The result is a scalable, secure, and modular pipeline that enables near real-time micro- and meso-learning analytics (\cite{santana2025}) and drives bi-directional personalization for teachers, learners, and researchers (\cite{park2025}).

In this chapter, we present the goals, design principles, and early applications of A4L2.0, demonstrating its role in advancing equitable, data-driven, and personalized education for adult learners.

\section{Redesign of The Architecture for AI-Augmented Learning (A4L)}

\subsection{Data Sources}
The leftmost block of the pipeline diagram in Figure 1, labeled Data Sources, outlines the primary data sources that feed into the Data Engine 2.0. These sources include: 

\begin{itemize}[noitemsep]
    \item \textit{Student Information Systems (SIS):} These systems contain structured administrative data such as student demographics, enrollment records, academic status, and course rosters. This information is essential for contextualizing learning behaviors and enabling personalized experiences across courses and institutions.
    \item \textit{Learning Management Systems (LMS):} The LMS captures detailed interaction logs, assignment submissions, assessments, and engagement data. This provides the behavioral backbone for both micro- and meso-level learning analytics.
    \item \textit{AI Tools (via LTI\textsuperscript{®}):} AI-powered educational tools like Jill Watson (\cite{kakar2024a}), SAMI (\cite{kakar2024sami}), and VERA (\cite{an2021cognitive,an2025}) generate fine-grained interaction or event data based on how learners engage with these AI tools. These tools use 1EdTech LTI (Learning Tools Interoperability\textsuperscript{®}) standards to connect with the LMS.
\end{itemize}

\begin{figure}[h]
\centering
\includegraphics[width=1\linewidth]{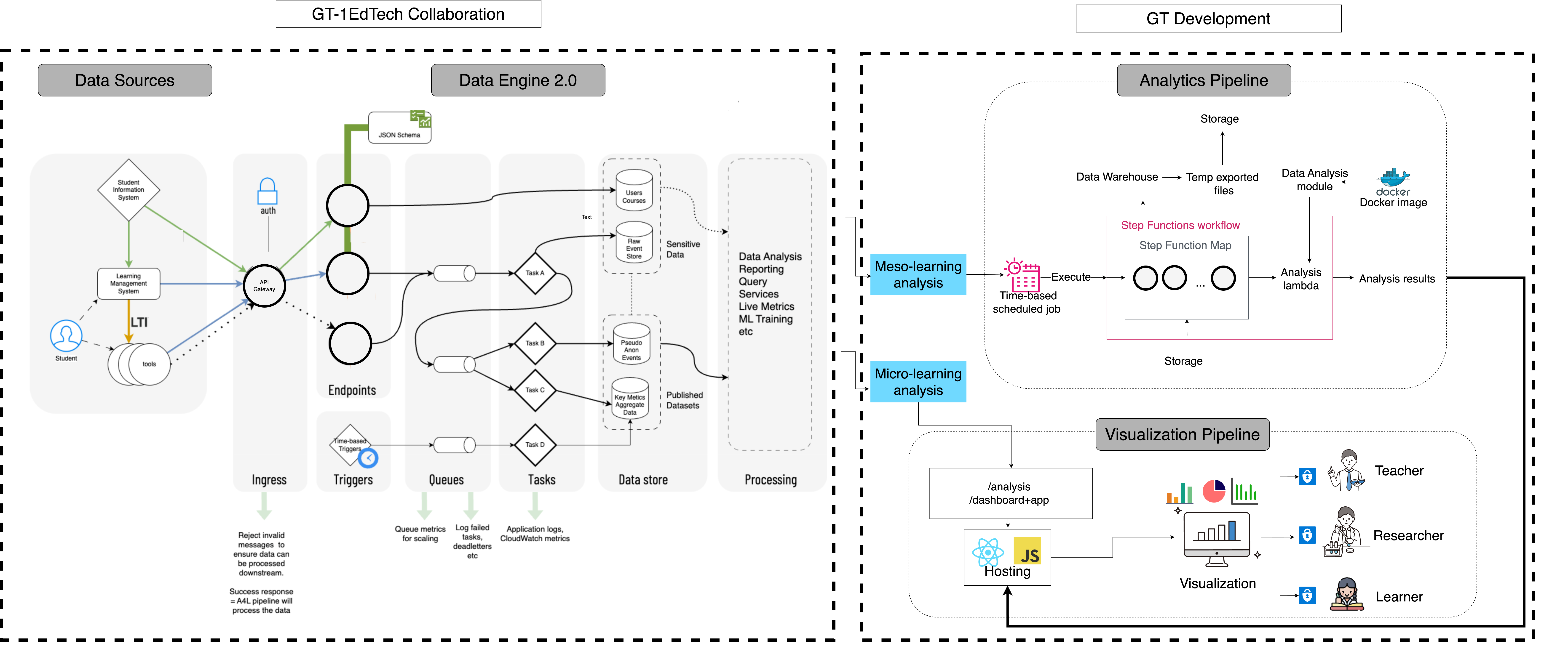}
\caption{\centering Conceptual architecture of the A4L (Architecture for AI-Augmented Learning) data infrastructure, illustrating the full end-to-end pipeline from data ingestion to visualization. The diagram consists of three main components - Data Engine 2.0 (left section of the figure): This section outlines the architecture’s data integration and processing backend. Educational data from Student Information Systems (SIS), Learning Management Systems (LMS), and AI tools (via LTI\textsuperscript{®}) are routed through a centralized API Gateway and validated against JSON schemas. These validated events are sent to system-specific endpoints, where time-based or event-driven triggers enqueue them for processing. The pipeline performs various preprocessing tasks. Analytics Pipeline (top-right section): This component executes scheduled meso- and micro-learning analyses. A time-based scheduler triggers an AWS Step Functions workflow that orchestrates containerized Lambda functions to extract data from the warehouse, transform it, and apply statistical procedures. The configuration is controlled via a declarative JSON payload, enabling modular and reproducible workflows for learning analytics and research. Visualization Pipeline (bottom-right section): The visualization layer delivers insights to end users—teachers, researchers, and learners—through interactive dashboards hosted using JavaScript (React). Analysis outputs are routed to specific applications such as Jill Watson, SAMI, or VERA dashboards, each tailored to user roles and needs. This layer supports near real-time monitoring, AI-generated insights, and secure, role-specific data access.}
\end{figure}

All these systems within Data Sources are integrated through 1EdTech open standards such as Edu-API, Caliper Analytics\textsuperscript{®}, and LTI\textsuperscript{®} which ensure secure, interoperable, and near real-time data exchange. Students interacting with these environments leave behind rich digital footprints, which are then ingested through an API gateway for preprocessing and analysis within the A4L pipeline.

\subsection{Data Engine 2.0}
The Data Engine 2.0 serves as the core of the A4L architecture, responsible for securely processing and organizing data from various educational systems before it is used for analytics and personalization. It consists of several key stages:

\subsubsection{Design Guidelines}
Based on earlier experiments with A4L1.0 (\cite{goel2025}), and to support scalable, secure, and maintainable data infrastructure, we created a set of design guidelines that specify a multi-environment architecture and modular data pipeline grounded in Amazon Web Services (AWS) and 1EdTech best practices. Data ingestion is extracted from diverse data sources with consistent identifiers aligned at the point of data ingestion (e.g., GT student ID, SSO) across tools like Jill Watson, SAMI and VERA. All data, including surveys, Canvas grades, and Ed Discussion posts, are centrally stored and organized by semester and course. Caliper Analytics® is used to manage schema transformation formats, ensuring consistency with 1EdTech metadata standards. The orchestration of workflows will leverage AWS services, while data will be staged and stored using S3 and Redshift.

Robust deployment and DevOps practices are critical for repeatability and extensibility. All Lambda functions are managed using Docker or Lambda layers with centralized dependency and image repositories. Logging and error handling are built into AWS CloudWatch and integrated with alert systems. A centralized template GitHub repository will serve as the foundation for testing, collaboration, compliance, and automation, ensuring scalable, maintainable, and secure infrastructure across A4L’s evolving analytics and visualization ecosystem.

\subsubsection{Ingress and Endpoints}
Data flows into the system through the API Gateway, which authenticates and routes data from 1EdTech’s data standards Edu-API (e.g., SIS and LMS records), Caliper Analytics\textsuperscript{®} (capturing learner events), and LTI\textsuperscript{®} tools (external AI tools). These inputs are passed to dedicated endpoints, including:

\begin{itemize}[noitemsep]
    \item EduAPI for roster and course data
    \item Caliper Endpoint for learning activity events
    \item Application Events Endpoint for tool-specific interactions
\end{itemize}

Each endpoint accepts data that conforms to a predefined JSON Schema, ensuring consistency and structural integrity.

\subsubsection{Triggers and Queues}
The architecture supports time-based triggers to schedule ingestion jobs and event-based triggers for real-time processing. Once data is accepted, it enters a queuing system that:

\begin{itemize}[noitemsep]
    \item Buffers incoming streams
    \item Enables scalability
    \item Handles data bursts or processing delays
\end{itemize}

These queues direct data to appropriate preprocessing tasks in an orderly and asynchronous manner.

\subsubsection{Preprocessing Tasks}
This layer provides a pattern for extending the transformation of raw data into structured, privacy-aware, and analysis-ready formats. For example, it includes:

\begin{itemize}[noitemsep]
    \item Task A: Stores raw, sensitive data (including user identifiers and course interactions) in the Raw Event Store.
    \item Task B: Applies pseudonymization techniques to create Pseudo-Anonymization Events, protecting user privacy while preserving analytical value.
    \item Task C: Aggregates data into Key Metrics (e.g., participation, engagement trends), which are used for dashboard summaries and learning insights.
    \item Task D: Can be used for auxiliary processing such as logging, error correction, or metadata enrichment.
\end{itemize}

Further tasks can be added to meet future needs without disrupting the existing flow of data.

\subsubsection{Data Store and Processing}
Processed data—both sensitive and pseudo-anonymized—is stored securely in a centralized data store. This repository supports downstream data utilization into our Analytics Pipeline and Visualization for personalized learning data analyses and visualization that includes instructor and learner feedback through a strict data governance structure.

The architecture ensures that all data is properly organized into published datasets, ready for use in research, near real-time analytics, or institutional decision-making. On the other hand, sensitive information is being stored in a separated location for privacy protection to mitigate against the risk of incorrect data access rules in the future, exposing sensitive data.

In summary, the Data Engine transforms multi-source educational data into standardized, privacy-protected, and analytics-ready formats—forming the foundation of the A4L pipeline’s ability to deliver personalized, scalable learning experiences.

\subsubsection{Testing}
We evaluate A4L2.0 using a comprehensive functional and non-functional testing strategy. The primary objective is to ensure that the pipeline meets both stakeholder expectations and technical requirements. The evaluation focuses on the core components of the pipeline—namely data ingestion, transformation, cleaning, and storage—while also validating the infrastructure configuration through infrastructure-as-code practices.

\subsection{Analytics Pipeline}
Learning analytics is increasingly central to the design and operation of contemporary educational systems, providing data-informed insights to guide both teaching practices and learner support. By examining behavioral data, engagement patterns, and academic outcomes, learning analytics supports personalized pathways, just-in-time interventions, and improved instructional strategies (\cite{siemens2011,ferguson2012}). Despite these opportunities, persistent challenges limit its effectiveness—particularly around data fragmentation and lack of interoperability. Learning data are often siloed across disparate systems such as LMS, SIS, and third-party tools, creating integration barriers and complicating longitudinal analyses (\cite{vidal2017}).

Several educational data platforms have been developed to address the challenges of data fragmentation and interoperability, including the Total Learning Architecture (TLA; \cite{schatz2019modernizing}, DataShop (\cite{koedinger2010datashop}), the Unizin Data Platform \footnote{\url{https://unizin.org/knowledge-base/}} , and My Learning Analytics \footnote{\url{https://its.umich.edu/academics-research/teaching-learning/my-learning-analytics/}} (MyLA). These platforms emphasize data aggregation and standardization but generally fall short in offering configurable analytic engines that enable flexible, replicable workflows and accessible cost structures for researchers and educators. Recent studies (\cite{donadio2021,sghir2023}) have underscored the growing need for modular, interoperable infrastructures that can adapt to evolving instructional demands and research contexts in real time.

To bridge this gap, the A4L 2.0 analytics pipeline - depicted in the top rightmost block of the overall pipeline in Figure 1, was developed. We introduce a data architecture specifically designed to support meso-learning—patterns of learner behavior that unfold across intermediate timescales such as a school’s semester. Meso-learning requires the integration of heterogeneous data sources and the application of Integrated Data Analysis (IDA) techniques (\cite{curran2018}) to surface trends that can inform timely pedagogical decisions. At the same time, \cite{lyndgaard2024} provide a conceptual framework defining meso-learning within a larger ecosystem of micro- and macro-learning processes. As we describe in \cite{santana2025}, A4L 2.0 meets these needs by framing the analytics pipeline as a Highly Configurable System (HCS) —an evolution of Software Product Lines (SPLs). This architectural approach enables runtime flexibility through the use of structured configuration files, allowing users to customize analytic workflows dynamically without modifying underlying code. HCS-based designs are particularly effective for supporting replicability and reuse across diverse learning environments, making them well-suited to the demands of real-time educational research and adaptive instruction.

The analytics pipeline is a modular and scalable system designed to process educational data for meso- and micro-learning analysis. This pipeline operates within a cloud environment and is built around AWS Step Functions to orchestrate the flow of tasks. It is triggered by a time-based scheduled job, which launches the analysis at regular intervals—such as daily or weekly—based on the desired granularity of insights. This scheduler initiates a Step Functions workflow that executes a predefined sequence of operations using Lambda functions encapsulated in a Docker container, ensuring modular, reproducible execution across environments.

The connection to the data engine occurs through the meso-learning analysis module. Data collected through the A4L Data engine 2.0—including structured event data from the LMS, SIS, and AI tools—is routed into preprocessed and standardized forms via key processing tasks such as anonymization and aggregation. Once transformed, the data is directed to the meso-learning analysis module, which acts as a bridge to the analytics pipeline. Here, relevant datasets are extracted from the data warehouse using a temporary storage mechanism (e.g., Parquet files), enabling the data analysis module to apply configurable statistical procedures. The results are then sent downstream for visualization or archival, supporting ongoing research and instructional interventions. This seamless integration between the data engine and analytics pipeline supports timely, scalable insights aligned with A4L’s goals for personalized learning.

In summary, our development introduces a modular, event-driven analytics pipeline that applies HCS principles to the domain of educational data. The system includes a configuration-driven engine for defining data sources, preprocessing steps, and analytic procedures. Implemented within a cloud-based, serverless infrastructure, and containerized for reproducibility, the pipeline integrates seamlessly with visualization platforms. Ultimately, we aim to demonstrate how this configurable analytics engine supports scalable, reproducible, and domain-specific learning analytics as a core component of the A4L ecosystem.

\begin{figure}[h]
\centering
\includegraphics[width=0.7\linewidth]{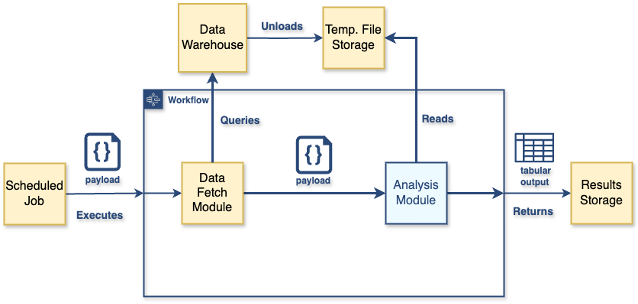}
\caption{\centering A high-level diagram of the data analytics pipeline. This diagram illustrates the core workflow of A4L's modular and configurable data analytics pipeline. The process begins with a time-based Scheduled Job, which triggers the pipeline by passing a JSON-based payload—a structured configuration file that defines parameters for the analysis. The Data Fetch Module uses these parameters to generate SQL queries to the Data Warehouse and initiates a data extraction process. The extracted data is unloaded into Temporary File Storage in a columnar format (e.g., Parquet) for efficient downstream processing. Once available, the Analysis Module retrieves the files and performs cleaning, transformation, and statistical analysis operations as dictated by the same payload. The analysis results—typically in tabular form—are stored in Results Storage, ready to be used by visualization dashboards or exported for reporting and further research. This modular pipeline supports reproducible, scalable, and configurable analytics within A4L's event-driven infrastructure }
\end{figure}

The architecture of the analytics pipeline is implemented as a state machine that is triggered by events in a cloud environment, as illustrated in Figure 2. In our deployment on Amazon Web Services (AWS), this trigger is defined as a time-based scheduled job that activates the pipeline at fixed intervals (e.g., daily at 9:00 AM). When triggered, the state machine begins an execution by consuming a structured input known as the payload, which contains the configuration and parameters for the analysis run.

Upon activation, the pipeline initiates the Data Fetch Module, which accesses the relevant datasets specified in the payload. This module interacts with a data warehouse by executing an UNLOAD command that exports data in a compressed, machine-readable format (Parquet) to a temporary cloud-based file store. The module then polls the warehouse at set intervals until it receives a “Complete” status along with file locations for each dataset. This metadata is passed forward for processing.

The Analysis Module—depicted in Figure 3 and as part of Figure 2—is the central logic component of the pipeline. It ingests the retrieved datasets, applies data cleaning and preprocessing steps based on functions defined in the payload, performs data transformations, and executes the appropriate statistical operations. The final analysis results are structured and written to a permanent cloud storage location, making them available to downstream applications such as dashboards or visualization tools.

\begin{figure}[h]
\centering
\includegraphics[width=0.7\linewidth]{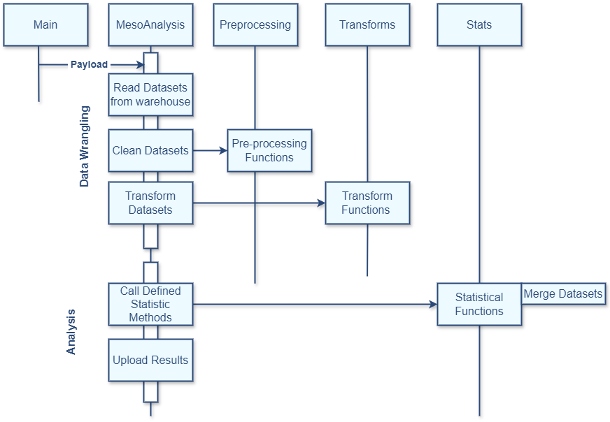}
\caption{\centering A sequence diagram of the analysis module. This figure illustrates the internal execution flow of the A4L analysis module, which is triggered by a configuration payload passed from the main application. The module begins with the MesoAnalysis script, which initiates the data wrangling phase by reading raw datasets from the data warehouse, cleaning them, and applying initial transformations needed for analysis. The cleaning stage invokes preprocessing functions, while transformation logic is handled by transform functions, both of which are modularized for reuse and adaptability. After data preparation, the pipeline proceeds to the analysis phase. The system calls specified statistical methods based on the payload parameters, which in turn reference external statistical functions defined in the Stats module. These functions perform operations such as aggregations, group comparisons, or regression modeling. Results from the statistical analysis may involve merging multiple datasets before finalization. The pipeline concludes by uploading processed results to a designated output location for downstream consumption in dashboards or reporting tools.}
\end{figure}

A key feature of this architecture is the payload, which acts as a declarative configuration interface. Encoded as a JSON file, it allows researchers to specify workflows in a modular and reusable way. In this implementation, a baseline payload was initially developed for a specific class in Fall 2023 semester and later generalized into a formalized template to support broader reuse. For analyses in the same domain, researchers can simply clone and modify an existing payload with minimal adjustments. The structure and semantics of this payload are captured in Figure 4, which defines the full configuration schema and allowable attributes.

\begin{figure}[h]
\centering
\includegraphics[width=0.7\linewidth]{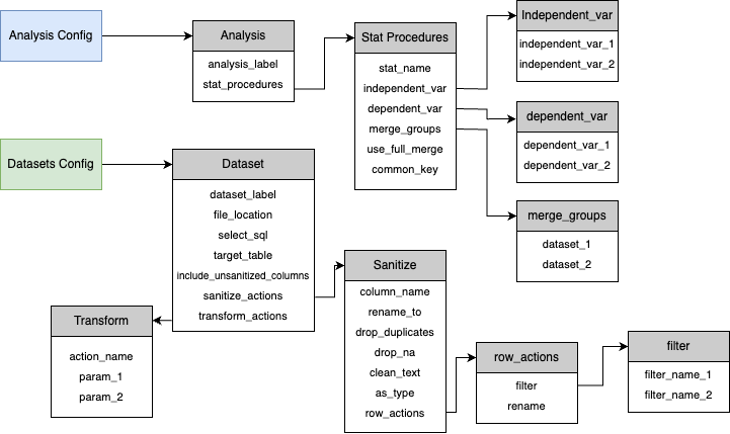}
\caption{\centering The payload’s data model. This figure illustrates the hierarchical structure of the payload configuration used to drive customizable analyses in the A4L analytics pipeline. The payload is composed of two primary inputs: the Analysis Config and the Datasets Config. The Analysis Config defines high-level analysis parameters, such as the analysis label and a list of statistical procedures to be executed. Each Stat Procedures block specifies a statistical method (e.g., regression, ANOVA) along with the variables of interest, including independent variables, dependent variables, and optional grouping logic defined in merge\_groups. It also contains metadata on how datasets should be merged using shared keys or full-merge logic. The Datasets Config outlines the data preparation workflow. The Dataset block specifies the data source (e.g., file location, SQL selection), the columns to include, and instructions for sanitize\_actions and transform\_actions. The Sanitize module includes row and column-level operations such as renaming, filtering, removing missing values, text cleaning, and data type conversions, some of which are further defined in row\_actions and filter blocks. The Transform module supports additional post-sanitization transformations with parameterized actions.}
\end{figure}

\subsection{Visualization Pipeline}
The visualization pipeline - depicted in the bottom rightmost block of the overall pipeline in Figure 1, is the architecture that serves as the final component for delivering insights to end users, including teachers, learners, and researchers. It is responsible for presenting analysis results in an accessible and interpretable format through dashboards and interactive visual tools. These visualizations are built and hosted using a JavaScript-based front-end framework such as React. The pipeline includes preconfigured applications that provide user-specific views of metrics, trends, and performance summaries derived from the underlying data analysis.

This visualization layer is connected to the data engine through micro-learning analysis and to the analytics pipeline through meso-learning analysis. After the data engine processes raw learning data and publishes cleaned, anonymized datasets, the micro-learning module extracts fine-grained, event-level data and routes it directly to the visualization layer. Simultaneously, the meso-learning module derives session-level or short-term behavioral insights and sends them to the analytics pipeline. The analytics pipeline performs targeted computations on these aggregated patterns and delivers the processed results to the visualization layer. The output is then rendered through web-based components that present actionable summaries of student behavior and instructional impact. This integrated flow supports near real-time feedback loops and enables data-informed decision-making by instructors and researchers within the A4L ecosystem.

\subsubsection{Pipeline architecture}

\begin{figure}[h]
\centering
\includegraphics[width=0.7\linewidth]{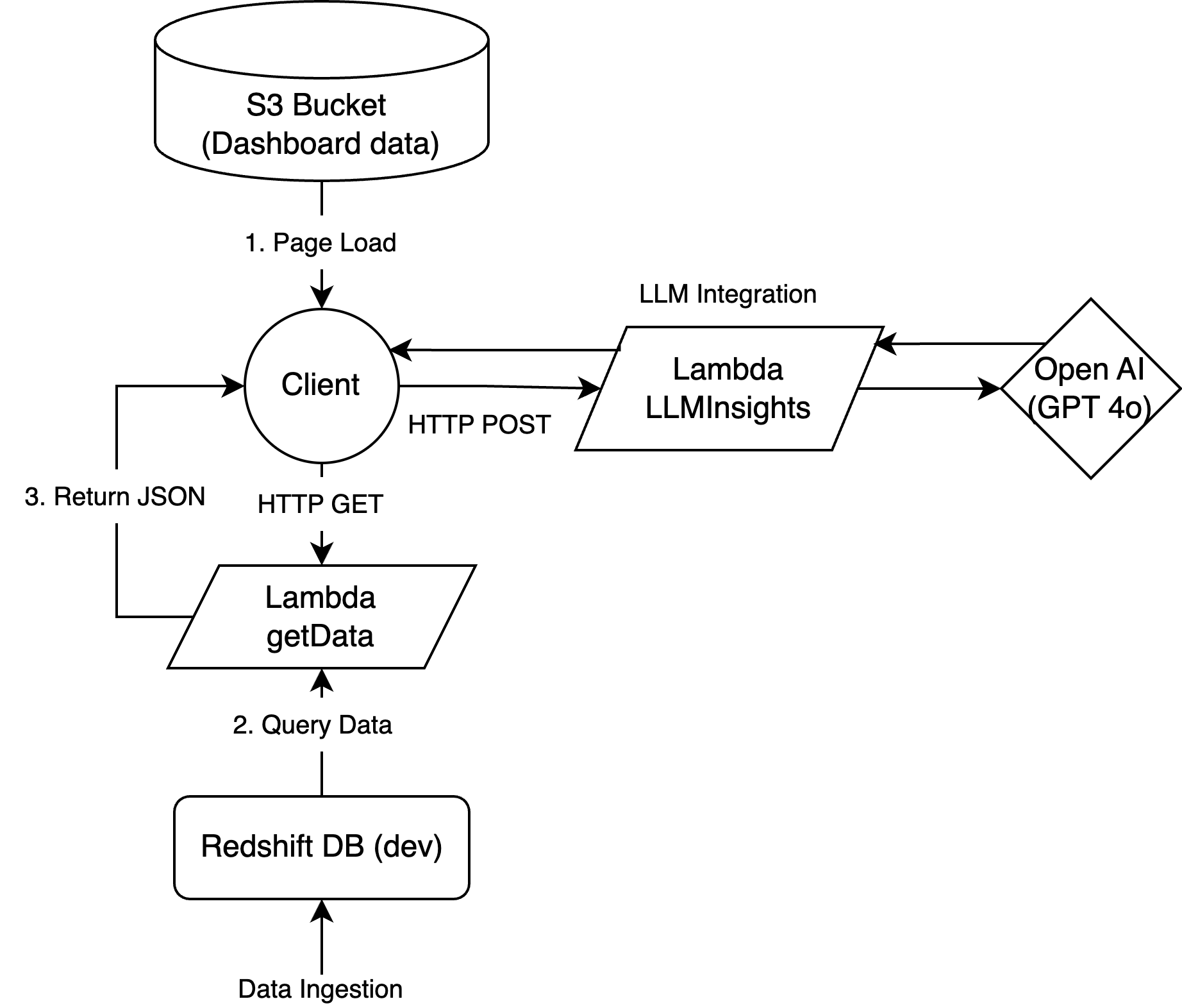}
\caption{\centering Overview of the visualization pipelines. This figure presents the high-level architecture designed to visualize insights derived from student interactions with AI tools. All dashboards support LLM-powered responses through integration with GPT-4o and follow a similar structure: a client initiates data queries, which are processed and returned for visualization. The Jill Watson pipeline (See Figure 6 below) features a fully automated data flow from source to dashboard, enabling near real-time updates. In contrast, dashboards from other AI tools (e.g., SAMI and VERA) currently rely on manually inserted data for querying and display. Despite differences in data ingestion workflows, all three dashboards utilize a shared architecture for insight generation and visualization, supporting personalized feedback and instructional decision-making across distinct AI learning tools }
\end{figure}

The visualization pipeline shown in Figure 5 above supports dashboards of three AI tools—Jill Watson, SAMI, and VERA—each following a similar architecture but differing in how data is ingested and stored.

The pipeline utilizes AWS Lambda to execute backend processing tasks and stores data in Amazon Redshift, a cloud-based data warehousing service, for efficient querying and analysis. Specifically, for the Jill Watson dashboard (Figure 6 below), the pipeline begins with automated data retrieval from the Jill Watson API using a Lambda function called JW-Pinq-API. This data is uploaded to an S3 bucket (jill-watson-data-store), which triggers a notification event. The event activates the JW-Queue-Event-Processor Lambda, pushing the task into a Simple Queue Service (SQS). Once the data reaches the queue, it is processed and loaded into a Redshift database (jw\_data\_current schema). A separate Lambda function, getJWData, queries Redshift to retrieve data and store it in another S3 bucket used for the front-end display. When a user accesses the dashboard, the client sends a request that triggers this data retrieval process. An additional Lambda function, JW\_LLMInsights, integrates with OpenAI GPT-4o to provide insight generation, which is then returned to the client for rendering.

\begin{figure}[h]
\centering
\includegraphics[width=0.5\linewidth]{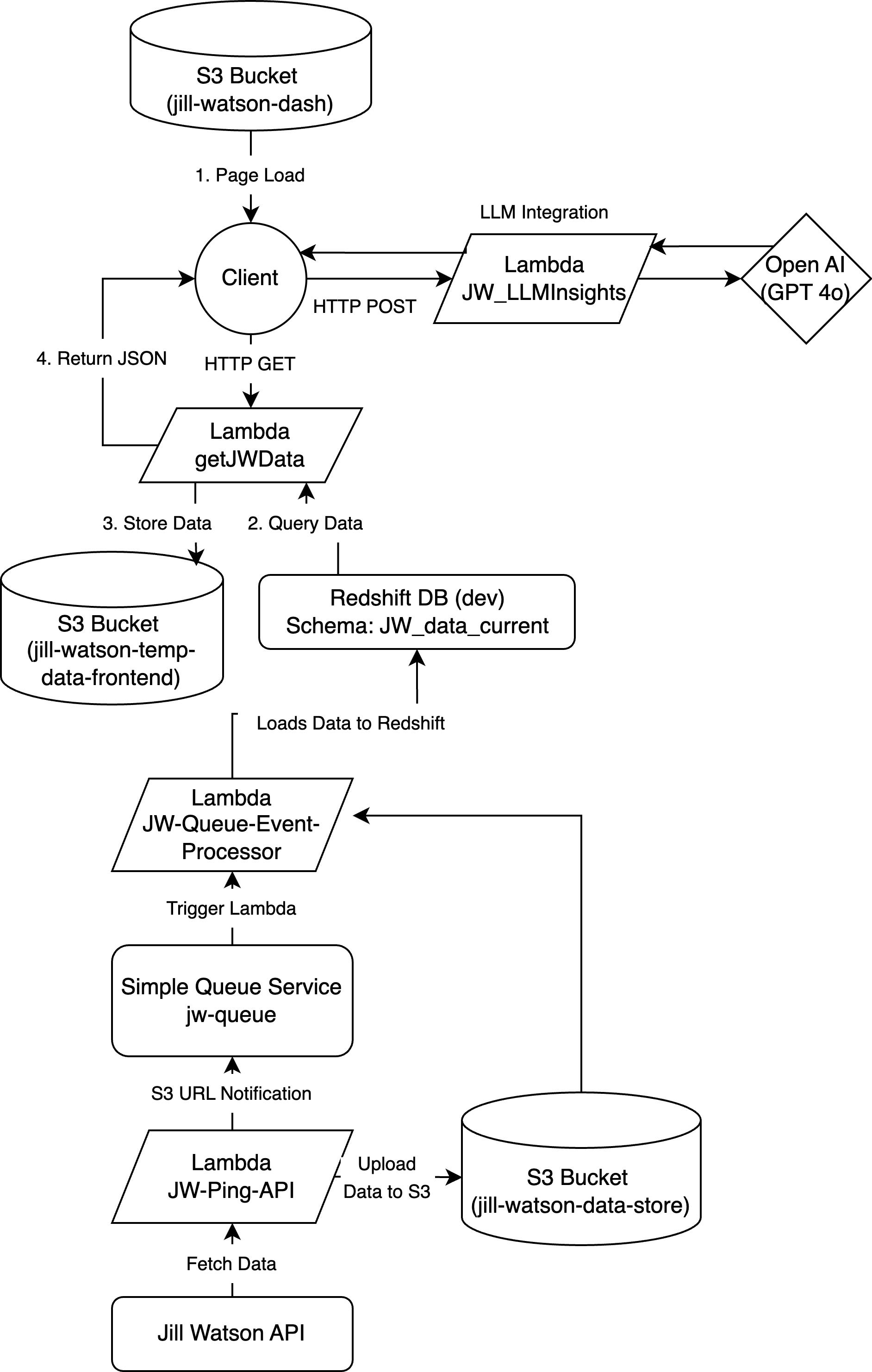}
\caption{\centering Jill Watson dashboard architecture for data processing and LLM insights. This figure illustrates the end-to-end pipeline for the Jill Watson (JW) dashboard. Data from the JW API is fetched by the JW-Ping-API Lambda, uploaded to an S3 bucket, and pushed to an SQS queue. The JW-Queue-Event-Processor Lambda then loads the data into Redshift (JW\_data\_current). When the client loads the dashboard, it queries data via getJWData, stores a local copy in S3, and returns the results as JSON. For AI-driven insights, the client sends data to the JW\_LLMInsights Lambda, which integrates with OpenAI GPT-4o to generate responses for visualization.}
\end{figure}

The SAMI dashboard follows a similar frontend pattern but differs in data ingestion. Data is manually inserted into Redshift. On page load, the client sends an HTTP GET request, which triggers the Lambda function to query the Redshift database and return the results as JSON. This data is then visualized in the dashboard. The Lambda function connects with GPT-4o to enrich the dashboard with language model-based summaries.

The VERA dashboard mirrors the SAMI workflow. Data is manually inserted into a Redshift schema. Upon loading the page, the client sends a request that invokes the Lambda function to query the database. The returned data is visualized in the dashboard.

The primary distinction across the dashboards lies in the data ingestion method—automated for Jill Watson and manual for SAMI and VERA—while all dashboards leverage LLM integration and Redshift for backend data processing.

\subsubsection{Example of Dashboard} 

\begin{figure}[h]
\centering
\includegraphics[width=0.9\linewidth]{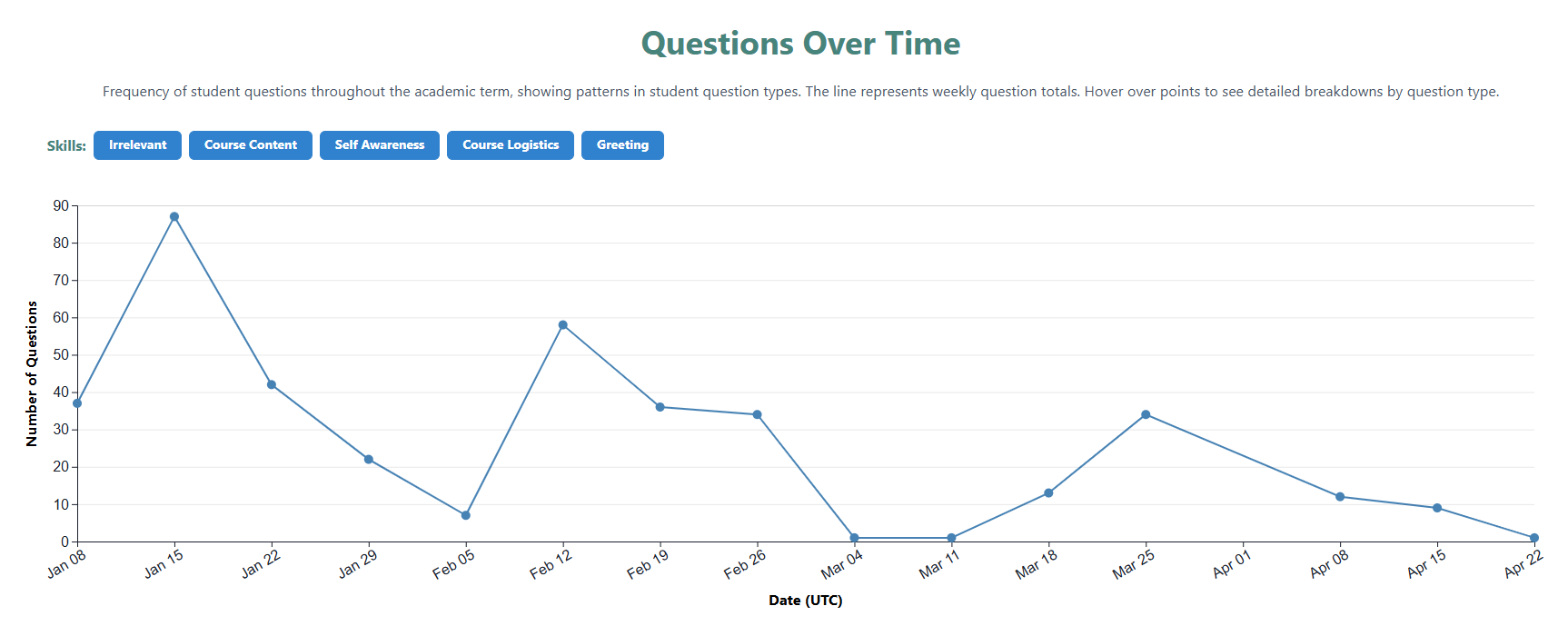}
\end{figure}

\begin{figure}[h]
\centering
\includegraphics[width=0.9\linewidth]{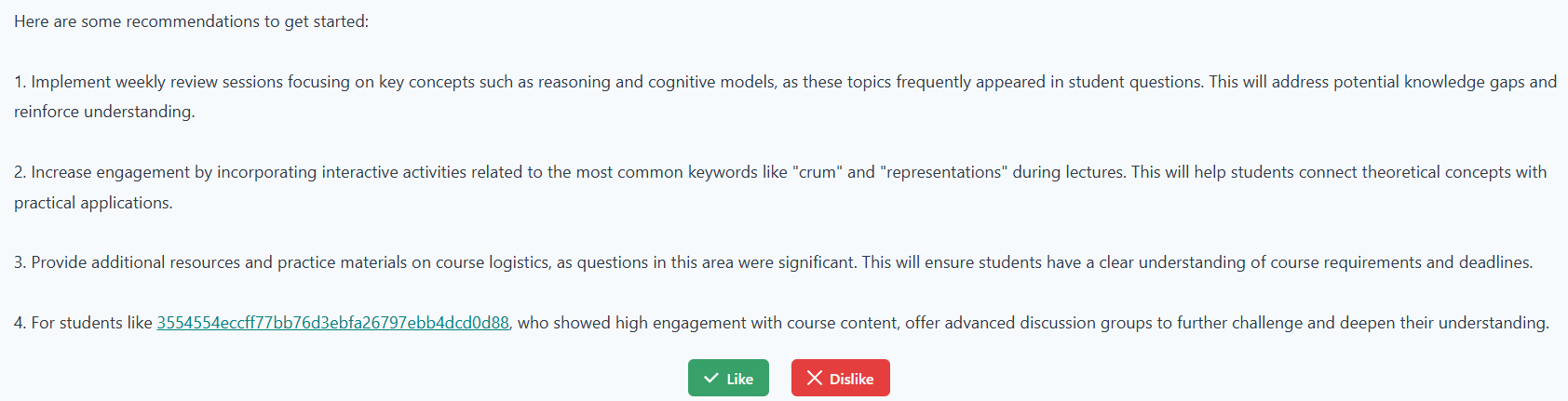}
\caption{\centering JW dashboard shows the “Questions Over Time” dashboard—a central feature of the A4L framework that supports the Bidirectional Feedback Loop between students, AI agents, and instructors. This dashboard visualizes the number of student questions over the academic term, offering a window into help-seeking behavior and learner engagement.}
\end{figure}

Integrated into a human-AI instructional loop, the data shown in Figure 7 is more than just a count of questions—it reflects the learner’s turn in the feedback cycle, where students interact with the system by asking questions. In the next step, the AI agent analyzes this interaction, generates insights, and updates the dashboard visualization. The instructor then interprets these visual cues, using their pedagogical expertise to adjust teaching strategies. These adaptations feed back into the system, allowing the AI to re-align its outputs, accordingly, completing the cycle (\cite{park2025}).

Peaks and troughs in question frequency highlight engagement shifts or possible confusion. For instance, a spike in early January suggests heightened activity or potential misunderstandings. Embedded recommendations below the graph offer targeted, AI-generated instructional actions—such as scheduling a meeting, sharing supplemental resources, or encouraging peer discussion—based on an individual student behavior.

This process is powered by embedding a LLM’s based AI agent that takes raw student data as input, contextualizes it, and generates actionable recommendations for the teacher. result is a dynamic, personalized dashboard that not only answers key instructor questions—Who needs help? When and where? What kind of help is needed? —but also delivers near real-time, interpretable, and personalized guidance to enhance teaching effectiveness.

In essence, this dashboard is both a visualization tool and a decision-support interface—enabling responsive, data-informed instruction through seamless human-AI collaboration.

\subsection{A4L for Human-AI Teaming}
A4L supports human-AI teaming by establishing a dynamic, collaborative feedback loop among learners, AI agents, and instructors (\cite{thajchayapong2025}). Rather than positioning AI as a standalone instructional tool, A4L emphasizes shared agency in which AI augments rather than replaces human judgment. As learners engage with digital platforms, their interactions generate continuous data streams that serve as the foundation for personalization. AI agents analyze this data in real time, generating predictive insights, adaptive recommendations, and timely responses to support the learner. These AI-driven insights are then interpreted by instructors, who apply their pedagogical expertise and contextual knowledge to refine instructional strategies, adjust course materials, and provide targeted support. A4L’s integrated dashboards and visualization pipeline enable educators to reflect on engagement trends and performance metrics, transforming complex data into actionable insights.

A central element of this human-AI teaming is the dynamic feedback loop that governs personalized learning environments (\cite{park2025}). This model highlights continuous and adaptive collaboration between the learner, instructor, and AI agent. At the center of this cycle is learner data—passively and actively generated through activities like completing assignments, asking questions, or interacting with digital tools—which serves as the foundational input reflecting each student’s progress, challenges, and evolving needs. The loop begins when a learner asks a question or seeks help, prompting an immediate AI response with answers or resources based on contextual and historical data. The instructor then interprets both the learner input and the AI output using broader analytics, refining instructional decisions. Updated data from the instructor is then factored into the AI’s support given to the learner, ensuring alignment with both pedagogical goals and individual needs. The loop culminates in the instructor providing personalized feedback, reinforcing key concepts or reshaping the learning experience. This iterative process creates a holistic, data-informed cycle of co-instruction. By placing the learner at the center and balancing automation with expert human input, A4L fosters more responsive, effective, and personalized learning experiences—demonstrating a robust framework for human-AI teaming.

\section{Conclusions}
The Architecture for AI-Augmented Learning 2.0 represents a comprehensive and modular solution to the challenges of personalization, scalability, and integration in data-driven online education. By aligning with open standards such as 1EdTech’s open data standards Edu-API, Caliper Analytics\textsuperscript{®}, and LTI\textsuperscript{®}, the A4L pipeline establishes a robust and secure foundation for collecting and transforming educational data across diverse platforms. Its fully asynchronous and event-driven design enables it to operate in near real time, ensuring continuous data flows from Student Information Systems (SIS), Learning Management Systems (LMS), and AI-powered tools like Jill Watson, SAMI, and VERA. These design principles support the growing need for flexible, privacy-aware, and reproducible analytic infrastructures that are crucial for modern educational ecosystems.

One of the central innovations of A4L is its layered architecture that connects the data engine, analytics pipeline, and visualization modules into one cohesive system. The data engine acts as the backbone for ingesting, preprocessing, and transforming raw interaction data into structured formats through well-defined tasks. These include pseudonymization, aggregation, and standardization—all of which are vital for ensuring privacy and consistency across analytics and visualization workflows. The pipeline supports both meso-learning and micro-learning use cases, providing longitudinal insights into student learning behaviors as well as near real-time dashboards for learners, educators and researchers. This dual capacity positions A4L as a foundational infrastructure for both formative feedback and summative evaluation.

The analytics pipeline, built using cloud-native orchestration tools like AWS Step Functions and Docker containers, further enhances the modularity and reproducibility of educational analysis. It adopts a declarative configuration interface—encoded in JSON payloads—that allows researchers and practitioners to define analytic procedures without modifying core code. This approach aligns with the principles of Highly Configurable Systems (HCS), which emphasize adaptability and maintainability in complex environments. Feature modeling and benchmark comparisons reveal that the pipeline maintains a favorable balance of flexibility and correctness, supporting over a billion valid configurations without introducing excessive structural complexity.

On the front end, the visualization pipeline completes the feedback loop by presenting insights in intuitive, role-specific dashboards for instructors, learners, and researchers. It translates micro-learning outputs into actionable visualizations and integrates with large language models like GPT-4o to support interpretability and contextualized feedback. The differentiated workflows for Jill Watson, SAMI, and VERA demonstrate how the architecture can support a range of educational tools and use cases, each with its own ingestion, storage, and rendering pathways. These dashboards not only improve visibility into learner behavior but also support iterative refinement of instructional strategies based on empirical evidence.

Looking ahead, A4L offers a compelling framework for addressing emerging challenges in online education and the awarding of micro credentials at scale. Future enhancements may include the incorporation of distributed computing for scalability, further decoupling of analytic components for modular extensibility, and a deeper integration with federated learning protocols to strengthen privacy guarantees. As educational institutions continue to adopt AI-driven technologies, architectures like A4L will play an essential role in bridging data interoperability, analytic rigor, and pedagogical impact—ultimately advancing a more equitable and personalized future for lifelong learners.

\section*{Acknowledgements}
This research has been supported by grants from NSF grants US National Science \#2247790 and \#2112532 to the National AI Institute for Adult Learning and Online Education (aialoe.org). We thank members of the A4L team for their contributions to this work including JunSoo Park, Matthew Patton, Natalia Theodora, Jayson Brown and Anthony Santana. 

\section*{Glossary}

\paragraph{API Gateway:} A service that acts as a single entry point for applications to access backend services. It manages request routing, authentication, monitoring, and scaling for APIs.

\paragraph{AWS S3 (Simple Storage Service):} A scalable object storage service. It is used to store and retrieve any amount of data (files, images, backups, logs, datasets) with high durability, availability, and security. Data is stored as objects within buckets.

\paragraph{AWS Redshift:} A cloud-based data warehouse service that allows for large-scale storage and analysis of structured and semi-structured data. It supports SQL queries and integrates with business intelligence (BI) and analytics tools, optimized for high-performance queries across petabytes of data.

\paragraph{AWS Lambda:} A serverless compute service that lets users run code without provisioning or managing servers. Code is executed in response to events (e.g., S3 uploads, API Gateway requests) and automatically scales based on demand.

\paragraph{AWS CloudWatch:} A monitoring and observability service for AWS resources and applications. It collects and tracks metrics, logs, and events to provide real-time insights into system performance, application health, and resource utilization.

\paragraph{AWS Step Functions:} A serverless orchestration service that coordinates multiple AWS services into workflows. It uses state machines to manage task execution, retries, error handling, and parallel processing, enabling complex automation pipelines.

\paragraph{Docker:} An open-source platform for developing, shipping, and running applications in lightweight, portable containers. Containers package applications and their dependencies together, ensuring consistency across development, testing, and production environments.

\paragraph{Highly Configurable System (HCS):} A software system designed with a high degree of flexibility, enabling users to configure features, options, and workflows without rewriting code. Often supported by techniques like feature modeling, parameterization, and modular architectures.

\paragraph{MongoDB:} A NoSQL database that stores data in flexible, JSON-like documents rather than relational tables. It is schema-less, supports high scalability, and is widely used for applications requiring fast, flexible data storage.

\paragraph{MySQL:} An open-source relational database management system (RDBMS) that uses structured query language (SQL).

\paragraph{Parquet Files:} A columnar storage file format optimized for analytics and big data processing. It provides efficient data compression and encoding.

\paragraph{Pull Requests:} a mechanism used in version control systems (such as GitHub, GitLab, or Bitbucket) to propose changes to a codebase.

\paragraph{Software Product Lines (SPL):} An approach in software engineering where a set of related software systems is developed from a shared set of core assets. SPL emphasizes reusability, variability management, and systematic configuration to reduce development time and cost while supporting customization.

\bibliographystyle{plainnat}
\bibliography{a4l20_refs}

\end{document}